# Expected participation and mentality of smart citizen in smart cities

Katalin Feher[0000-0003-3293-0862]

Budapest Business School University of Applied Sciences, Budapest, Hungary

**Abstract**
The purpose is to investigate the expected participation and mentality of smart citizens in smart cities. The key question is the role of the human factor in smart environments globally studied through a research corpus of 150 documents including mainstream summaries trend reports, white papers and visions of business – governmental – university research co-operations reaching a wide audience of the subject. Foremost, a short review of the changing scholarly trends is presented as a theoretical framework. Concerning its key ideas, the corpus based findings are recapped and analysed by content networks and the most referred city strategies. Besides, a critical approach reveal further required factors and risks to investigate. The ultimate goal is to understand how the smart city landscape is shaped by citizen-based strategies, open data, empowerment and responsibility. Accordingly, the paper closes with further considerations regarding the importance of anonymous open data, advantages of neighbourhood-based implementations, aspects of permanent and temporary citizen-engagements, interpretation of metaphors or upcoming technologies, and also, privacy and ethical issues. The results provide the policy development and the emerging scholarly interest with a framework study.

**Highlights**
- Summarises the changing scholarly trends regarding "smart city and smart citizen"
- Presents governmental-business-university co-operations based on 150 sources
- Analyses the participation and mentality via subject nodes and highly cited cities
- Explores the low-represented issues in the corpus from privacy and ethics to AI
- Interprets metaphors, potentials and risks in the socio-technical engagement

## 1. Introduction

Smart city is trending now in governmental, business strategies and in academic research not only with adjective "smart", but also, with further definitive prefixes "cognitive" and "intelligent" (among others Ishida, 2017; D'Onofrio and Portmann, 2017). The "smart" category has already referred to certain human sensing and control regarding digital operation. Alternatively, categories of "cognitive" and "intelligent" presuppose more complex mental operation which probably become comparable to the human substances as constant reference points in terms of philosophy.

In parallel to the extensions of artificial creations, the attention has been increasing to the additional human participation and mentality in smart or intelligent environments. The reason behind this growing interest is that the human existence has been less and less available to investigate independently of the current technology which feeds on data sets of general public and its surrounding services together. Consequently, the human factor has become less insightful in itself. It is therefore suggested to identify the expected contemporary expected human participation and the adaptive mentality while it is available.

Paradoxically, the results may improve the non-human substances, which is not a subject of this study.

One of the most pervasive digital environments, namely smart cities, allow a complex investigation of human factors on the above mentioned field. Smart cities assume not only a digitalised infrastructure with intelligent service network and data-driven decisions (Feher 2018), but also, an ongoing reflection flow of citizens, communities or any kind of other human participation. Moreover, reflections and adaptations of users are also shaping the digital developments. This interactive connection between technology and general public results a field to be explored for interpretation of smart city landscape.

Additionally, a significant approach, smartmentality (Vanolo, 2014) presumes a conscious and mature usage of digital technology. Although Vanolo discussed this term concerning responsibility of policy making and political-financial decisions, this term might be extended also to smart citizens and their groups. Their mindset, attitude and adaptive behaviour play a pervasive role in responsible activities, engagements or experiments. This interpretation is also confirmed by relevance of liveability, quality of life, predictability, comfort and human values (Gudowsky *et al*, 2017; Hernafi *et al*, 2016). Consequently, efforts of citizens to apply the smart technology properly and the holistic goal of the well-being in the cities reflect each other.

Along these paths, the goal of this paper is to investigate the expected participation or mentality of individual and collective users in smart city. To study the contemporary strategies and practice of business-governmental-research co-operations on this field, a research corpus has been created via mainstream executive summaries, trend reports, white papers and visions. The paper formulates the specific human aspects of digitally defined urban environments presenting topic networks of the corpus and a summary of the most cited city cases. The ultimate goal is to understand, how the current and future city landscapes are shaped by human-centric strategies, and also, to support the business and political decision making by a global perspective.

## 2. Changing trends in the literature. A short review

According to the scholarly data bases, the number of academic papers about smart cities has been growing rapidly in the last decade (Lim *et al*, 2018). Investigating only a few major public and restricted databases (from Google Scholar to Science Direct, EBSCO or JSTOR), the intensive interest started to expand around 2011 and the growth is about between ten to twenty times in average until the end of 2017.

Smart services and their converging developments have become only one of the reasons behind this trend (Soto *et al*, 2015). The promise of artificial intelligence (AI) and the introduced artificial narrow intelligence (ANI) (among others Burgess, 2018) with limited cognitive or intelligent services have been anticipating a structural change integrating IoT, cloud computing and robotics. The concepts of smart cities have been fundamentally studied or criticised as the emphasised technology in themselves (Han & Hawken, 2018; Greenfield, 2013).

Besides, the upcoming paradigm shift via new digital technologies, such as autonomous vehicles or optimised public utilities, facilitates a changing view of governmental and business strategy in the cities, especially regarding on institutional changes and adaptation attitudes by citizens (among others Meijer and Bolívar, 2016). Therefore, the massive non-human focus from AI to robotics also promotes the human factors to be investigated more deeply. Concurring with the idea of Gudowsky and his co-authors (2017), societal needs, demand-side thinking, inhabitant participation in policy economy provide a contemporary perspective in smart city strategies. In close association with this, it is also considerable to rethink the social infrastructure (Han and Hawken, 2018) beyond the above mentioned predominance of technological view. Although, these approaches are presenting only a few issues, otherwise, highlighting the most though-provoking and future-oriented ideas of the human factors in smart city context. The goal of this paper to contribute to this discourse.

Considering the forthcoming technological impacts, the specific part of the city strategies should include the citizens with their data sets and digital footprints (Feher 2017) via online or sensorised systems, and also, their activities and engagement to smart environments. Although this view has been still less pronounced compared to the technological view, its significance has been increasing (see the analysed corpus below). Scholarly sources have been following this slow movement (among others Benoit and Hiroko, 2016; Thomas *et al*, 2016). It is noticeable that the number of academic papers about the role of smart citizens is also rising intensively in the above mentioned databases. In contrast, this research field presents small number of academic papers compared to the wide infrastructural-technological approaches or the top down decision making issues.

Beyond the criticism of publication numbers, conceptual considerations are also suggested. Two specific academic sources are significant in arguing the fundamental problems with smart city definitions which results in non-proper interpretations of the smart citizenship. First, Vanolo (2016) emphasises how smart urbanism is a poor concept because of ambivalent visions about the role of citizens with different level of freedom and activities. In this context, wide spectrum of subjects, such as privacy or freedom of speech, are less significant in smart city strategies (Vanolo referring for Isin and Ruppert, 2015). According to his conclusion, a "public agora" would be expected with responsibility of citizens. Additionally, Neirotti and his co-authors (2014) emphasised that without a universal definition of smart city there is no available cornerstone for smart citizens. Otherwise, only the local context and the financial resources shape the characteristics of a smart city. Considering his approach, cultural-social-economic framework determine the smart citizens in different cities in different ways. These two papers are triggering a suggestion to choose in fact the smart citizens as cornerstones instead of looking for further non-comprehensive smart city definitions.

Concluding the literature review, a tipping point was detected compared to the infrastructural or technological academic sources (among others Barns *et al*, 2017; Phdungsilp, 2011) moving forward a growing attention to human factors (among others Gudowsky *et al*, 2017) and to the citizen-based strategies. In context of the upcoming human and non-human structural changes, simplified smart city definitions or only top-down concepts are not sufficient any more. Nevertheless, citizen-based or systematic cultural-social-economic approach improves the future city strategies.

Therefore, it is necessary to map globally and systematically the landscape of smart city and smart citizen approaches together to understand the above mentioned issues. Based on this consequence, the transdisciplinarity (Gudowsky *et al*, 2017; Brown, 2015) and a deep analysis of relevant documents would be supported to have recommendations for strategies of future cities. The following analysis of a globally filtered corpus has been developed, inter alia, for this purpose.

### 3. Corpus-based methodological concerns and the research questions

Following the original goal, investigation of expected participation and mentality in smart city concepts has become necessary in global perspective. A corpus was built to analyse the current trends and to understand the expected participation in smart cities. Smartmentality was also in the focus from the role of the citizen or community engagement to the empowerment.

Based on these considerations, the research questions were the followings:
1) What is the expected participation in smart cities by smart citizens in co-operations of governmental-business-university research?
2) What kind of smartmentality is supposed to be in a smart city by mainstream documents of governmental-business-university research collaborations?

To answer these questions, the above mentioned aggregation was studied to map the content networks and city strategies.

First and foremost, the corpus was selected and built by the most downloaded and linked documents of smart cities, based on the cumulative data of Google hits from three years (2014 Q2 – 2017 Q2). First, the keywords were "smart city", "smart citizen", "government*", "business", "university" and "research" applying together. The criteria was to find the first 150 hits of mainstream executive summaries, trend reports, project analysis, white papers and visions of governmental-business-university research co-operations from the millions of hits (G2B2UR as a 3D approach). The goal was to hunt the most viewed and popular hits to provide the answers for research questions with well-focused documents. These documents were available on the first pages of the search engine with significant visibility to reach wide audience. In this selection, 50-50-50 files were related to the sub-corpus criteria, identifying the main owners of the projects as governments, companies and universities.

Solely, English documents were filtered from the most clicked and cross-linked results for the subservient decision about the methodology of text analysis. A comparative analysis was available this way. Only the completed and visualised PDF versions of conceptions have been downloaded. Theses brochures and summaries warrantied a repeatable research compared to the ongoing-edited contents online. Moreover, the documents supposed wider audience due to the easy-understandable visual illustrations. In line with this, the dynamically changing web contents, such as summaries on websites or updated top lists, were not the part of the analysis intentionally.

It was crucial to define that the examined university research projects appeared only as a part of governmental and business documents. In other words, university research in this sub-corpora was not equivalent to a literature review. The reason behind this decision was to separate the university research projects in co-operation with governments and business from the academic discussion above. The audience of the selected documents in the corpus were supported to be from the potential business partners and the general public, while the interest of the conceptual academic sources belonged to scholars.

The key documents of the last three years were selected which means a short term. The advantage of this term was to reach the most updated strategic summaries of the rapidly changing technological issues. The disadvantage was that the corpus did not handle the proportions of long-standing discourses. This disadvantage was compensated by cited reference works in the examined documents.

The corpus was composed from one hundred and fifty documents counting in the 3D approach from last three years. Executive summaries, trend reports and collaborative projects for wider international publicity, summarised contemporary strategies and concepts with strong focus on expected participation.

After the corpus has formed, diverse methodology of conventional text statistics and text network analysis with data visualisation were applied. The networked data visualisations were built by the co-occurrence matrix of the text. In order to interpret the structural attributes and topics of the text, network metrics were applied. The network visualisation supported the qualitative interpretation of the data. Quadratic Assignment (QAP) Pearson correlation supplied the well-known correlation coefficient as a metric of comprehension for two matrices. Regarding the word pairs, two words were connected if they co-occurred in the same sentence and not more than three words were far from each other. Based on these terms, WORDij (Danowski, 2013) software was applied for the analysis. In the process of data visualisation and network calculation, the software Gephi 0.9.2 (Bastian *et al*, 2009) was utilised. In order to filter non-informative and conjunctive words from the corpus a stop list was applied. Connections were only counted if the two words co-occured at least two times. Connections based on word co-occurrences in text had no directions, therefore, the results of the network analysis were interpreted undirected ones.

On the network figures words were indicated as nodes, their size was equal to the amount of their importance. The importance of words was equal to their betweenness centrality (Brandes, 2001; Paranyushkin, 2011). Therefore, the sizes of the nodes were representing the amount of occasions one has to touch the node to connect two randomly chosen nodes with the shortest path in the network. Louvain modularity was applied in order to detect topic clusters in the corpus. Modularity algorithm (Newman, 2010; Fortunato, 2010) identified the communities within the network. Nodes ordered in the same community had more connections than it was expected on the basis of chance in a random network with the same amount of nodes and density. The coefficient of modularity equaled to the number of edges within a group of nodes minus the number of edges of the group of nodes in the random network. Gephi software used Louvain modularity with the standard 1.00 resolution (Blondel *et al*, 2008). Resolution was applied to encounter the fragmentation of poorly connected large networks. Transparency and qualitative interpretation was essential in data visualisation, therefore, an animation called Force Atlas (Jacomy, 2009) was chosen to render the networks with the most central nodes in the centre surrounded by connected nodes with smaller centrality from the same community.

Before this research project, a previous one investigated the most common and general key data-driven and infrastructural issues of contemporary smart cities. It was presented and published by SMART 2018: The Seventh International Conference on Smart Cities, Systems, Devices and Technologies. The paper won publication award. After the holistic approach, the next focus was on the expected citizen-dimension. The details are available below.

## 4. Findings I. Keyword and network statistics

The completed corpus was broken down into three subcorpora as the owners of the co-operations in 3D research. The three subcorpora were "business", "governance" and "university research".

Regarding the research questions, the purpose was to find the citizen-based expectations and smartmentality aspects in the documents. The analytical units of the corpus were selected as human factors based on the research questions. These parts have been found manually as protocol of text analysis (Krippendorf 2018). Having the key parts and elements, keyword statistics and word pair analysis were applied. The first part of the findings presents the connected top keywords and word pairs in the corpus providing by all documents. This part points out the correlation among the subcorpora as text networks too.

### 4.1. Keyword and word pair frequencies

Starting with the business-focused part of the corpus, the top ten keywords as frequency are connected to the human aspects from "people" or "citizen" to "communities". The framework is represented by open data, social, and also, private issues. Activities or engagement factors belong to the living-collaboration-participation pivot. The government-driven top words of the corpus highlights the "public" interest of activities by citizens and communities. The role of open data, participation and collaboration have also strong focus in this context. Public and private sector are also on the top list primarily reflecting services, authorities, stakeholders and safety. Last but not least, university research joins the topics of living issues, social perspective and collective activities of business and governmental projects.

Cumulating the results of the whole corpus, active participation forms are assumed by individual and collective users primarily by their data sets, collaboration and living issues. Comparing the differences, the business subcorpus highlights the collaboration aspects, the governmental subcorpus focuses on participation, and university research underlines the collective outputs. Private matters are more focused by governments and business (see tables 1-3. below).

Obviously, the occurrence of the keywords reflect cultural, economic or social contexts resulting in not exactly the same meaning of the words. Though, the goal and the size limit of this paper do not support to specify all factors with a sophisticated analysis, the section "Participation types in the top ten cities" below is going to present a few aspects of cultural, economic and social contexts.

| label | degree centrality | betweeness centrality | modularity_class |
| --- | --- | --- | --- |
| people | 16 | 3.927453 | 0 |
| living | 14 | 0.652778 | 0 |
| citizen | 16 | 1.742532 | 0 |
| collaboration | 15 | 1.323088 | 1 |
| open | 17 | 4.925866 | 0 |
| community | 16 | 8.387374 | 0 |
| participation | 12 | 0 | 0 |
| data | 16 | 3.570707 | 0 |
| private | 16 | 3.570707 | 1 |
| social | 16 | 1.742532 | 1 |

Table. 1. Network statistics of Business subcorpus - TOP 10 keywords

Table. 2. Network statistics of Government subcorpus - TOP 10 keywords

| label | degree centrality | betweeness centrality | modularity_class |
| --- | --- | --- | --- |
| people | 18 | 9.135365 | 0 |
| public | 15 | 1.053114 | 0 |
| citizens | 17 | 3.968698 | 0 |
| data | 17 | 3.968698 | 1 |
| private | 14 | 0.160256 | 0 |
| collaboration | 15 | 0.532984 | 0 |
| living | 15 | 0.532984 | 0 |
| community | 17 | 0.160256 | 0 |
| participation | 17 | 3.968698 | 1 |
| open | 15 | 0.532298 | 1 |

Table. 3. Network statistics of University-research subcorpus - TOP 10 keywords

| label | degree centrality | betweeness centrality | modularity_class |
|---|---|---|---|
| social | 17 | 3.963167 | 0 |
| living | 15 | 1.654401 | 0 |
| data | 17 | 3.963167 | 0 |
| public | 15 | 1.259199 | 1 |
| people | 17 | 3.963167 | 0 |
| open | 15 | 1.042929 | 0 |
| community | 17 | 3.963167 | 0 |
| participation | 18 | 11.296501 | 0 |
| collective | 13 | 3.333333 | 0 |
| communities | 15 | 4.681818 | 0 |

The further keywords in the top twenty are repeated the plural or singular versions of the above mentioned key terms, and also, synonyms and similar categories are available. The category of "empowerment", however, is also highlighted in the top twenty which is a remarkable result. It means a special participation type with additional authority, power and responsibility of citizens in decision making. This factor will be detailed in the next section.

In order to compare the content networks, the similarity of every two datasets has measured with the QAP Pearson correlation coefficient. QAP Pearson correlation compare two datasets with a standard number of 100 permutations expressing the similarity with the well-known correlation coefficient. Table 4 shows the result of the analysis as a correlation matrix.

Studying the whole corpus via correlation matrix (see Table 4), linear dependence is the strongest in case of government-business co-operations. Otherwise, university research presents a relevant and integrated role in business strategies, and also, in governmental policies. Obviously, this result partly derives from the corpus selection. By comparison, the strong linear dependence also confirms the importance of citizens in all three kind of strategies and reports substantially.

Table 4. Correlation matrix of Business-, Governance- and University subcorpora

|  | Business | University-research | Government | Number of random permutations |
|---|---|---|---|---|
| Business | 0.000 | 0.887 | 0.955 | 100 |
| University-research | 0.887 | 0.000 | 0.876 | 100 |
| Government | 0.955 | 0.876 | 0.000 | 100 |

Interpreting the top word pairs in the whole corpus, the above mentioned word lists are combined. The keyword "public" appears in different pairs of the whole corpus with the highest frequency. The wordpair primarily is the "private" regarding public or private sectors, and also, open data dilemmas. Comparing the top listed keyword frequencies, the additional factor is the the private aspect expanding the possible interpretations. The result is underlining a significant correlation between private and public based on open and non-open data distinctions. The case study section will also reflect this result.

**4.2. Text networks**

The above filtered terms and their strongly connected keywords provide a deeper analysis investigating related and further nodes in the text networks. In other words, a different methodology was applied in text network analysis compared to the keyword statistics which has resulted overlapped but partly different outputs. Keeping the 3D corpus, the contents of business, governmental and university research reveal the connections and average path of the keywords.

In the business-focused part of the corpus, there are 127 edges between the keywords. In average a node has 13.368 connections. This number is relatively close to the maximum number of the connections a node has (18), therefore specific combinations of the keywords are frequently co-occurring in the corpus. In average 1.26 steps need to be done to connect two keywords in the network. The longest distance between two nodes is 3 steps which presents a multiply strong connected content network. Nevertheless, the average path length indicates that certain keywords are presented as hubs, therefore, in average the components of network could be shortly connected. The density of the network measures 0.743. 74.3% of the possible connections between all of the nodes are presented in the network. According to modularity, there are 2 communities detected in the groups of keywords and these groups are more connected than it would be expected on the basis of chance (see Figure 1).

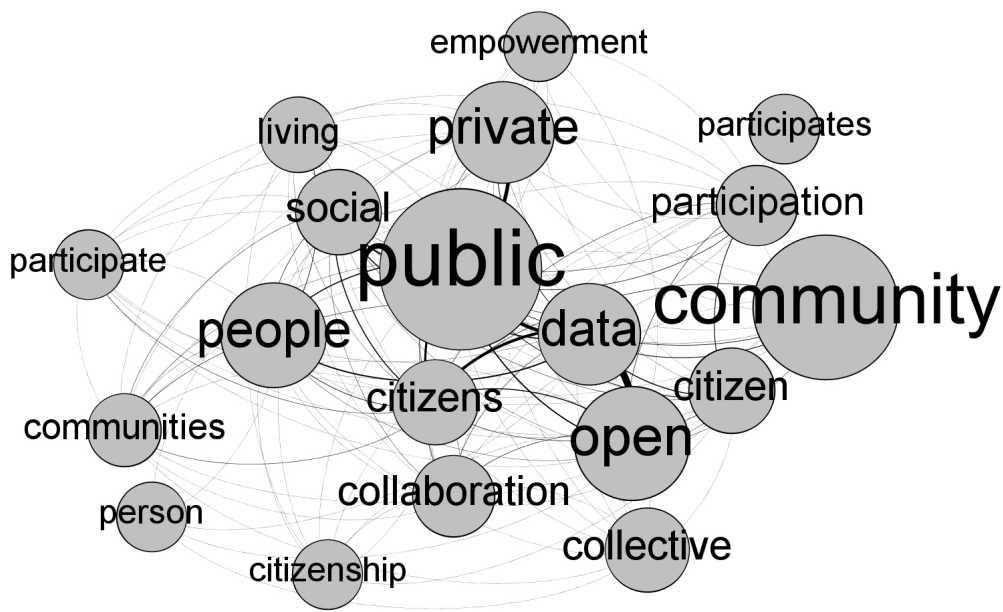

*Figure 1.* Text network of business subcorpus

The results of network analysis confirm the relevance of human factors in smart city projects. In the business subcorpus, the category of "community" is a hub instead of individuals such as "people" or "citizens". This result possibly derives from the parallel spotlight on "public" sector which is also a hub in this network. While people as individuals and citizens supposed to have private issues or privacy matters substantially, the public factors with communities assume collective participation, open data and collaborative willingness. The expected output is a contribution to open data and to a voluntary active operation supporting business goals. The question is how the category of "empowerment" fits into this logic from the periphery of the text network even via the private issues. With deep content analysis it was revealed that the meaning of "empowerment" in this subcorpus roots in engagement and associated with independent decision making done by citizens or participants of the local business. Illustrated with highlighted cases, empowerment facilitates innovations of micro-business, peer-to-peer platforms of aware customers, developments of collaborative models or investments of infrastructure. Consequently, the communities and the collaborations are in the focus in the context of open data and empowerment. Empowerment also reflects on the smartmentality with sophisticated types of possible engagement and triggered factors of bottom up developments.

The governmental subcorpus, compared to the afore mentioned result, has a bit more (130) edges. This corpus has almost the same maximum number of 18 and quite similar average degree which is 13.684. The average number of steps what has to be in order to connect two random nodes in the networks is 1.24. Therefore, the same attributes are presented as afore. The number of 2 longest path in the network indicates that this one has a bit shorter longest distance. The same conclusion is presented in density which measures 76% of the possible connections. A low positive coefficient of modularity indicates that groups of nodes are more likely connected that it would be expected on the basis of chance in a random network (see Figure 2).

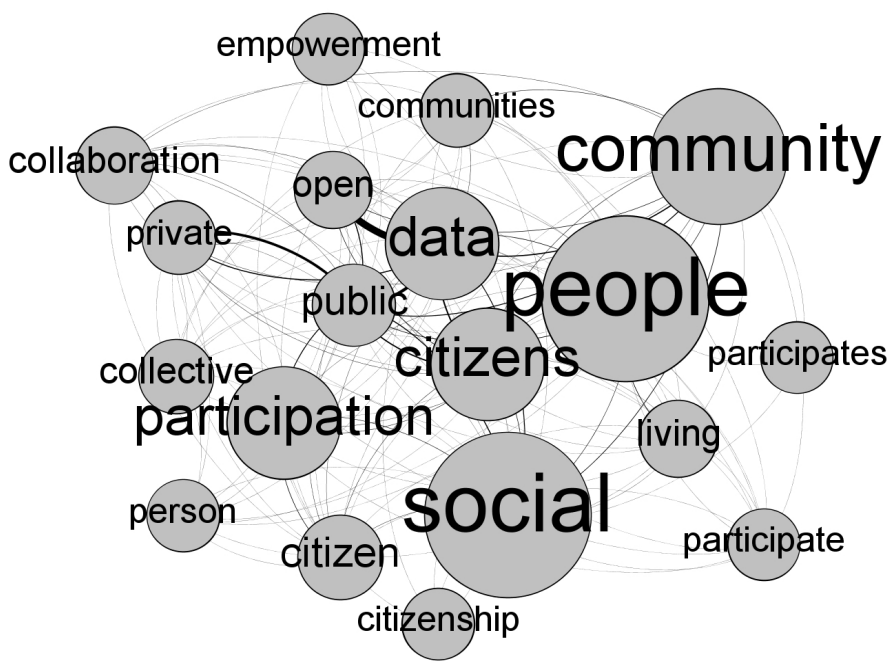

*Figure* 2. Text network of governmental subcorpus

According to the network analysis, the category of "people" forms a hub in the centre with intensively connected further central nodes, such as the correlated "citizens" and their public or open data, and also, the partly collective and partly personal "participation". Compared to the top ten keyword statistics, "social" factor is more connected and highlighted than it would be expected. The reason behind of this result should be the context of the society regarding the governmental view. "Collaboration" is on the edge assuming the role of "communities" behind, just like the "living" factors. The empowerment is also on the margin of the network compared to business subcorpus. However, the content analysis revealed different focal points to this expectation, such as philanthropic civic experiments, community leadership in daily lives, co-design at public places, transparency of open data to shape the city policy. "Social" issues, "people" and "community" shape a dividing border between two possible participation types. The first belongs to the "living" as geographical concerns and inhabitancy. The second extends the participation for digital and smart environment by "data", "collaboration" and "empowerment". Consequently, the physical and the virtual scenes have been merged in governmental strategies.

The text network, created from closely related university research of the corpus, has the same maximum number of degree (18) as the corpus mentioned afore as well as quite the quite close average degree which is 13.688. Therefore, specific combinations of words are frequent in the corpus. The 1.26 average path length indicates the option of large hubs in the network as well as the 2 diameter indicates words positioned away from these central hubs. The whole network is equally interconnected as afore due to the 0.743 density. Modularity has detected two communities which interconnectedness is close to what would be expected in a randomly interconnected network (see Figure 3).

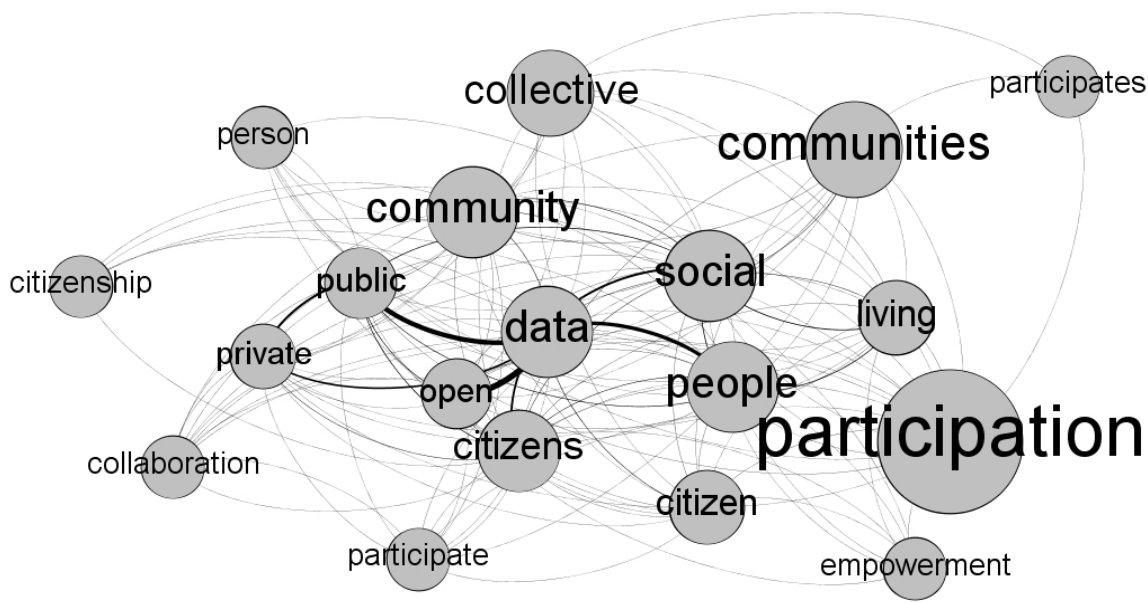

*Figure 3*. Text network of university research subcorpus

Although, the spotlight of "collective" matters were detected by the keyword statistics above, the text network resulted in the "participation" as a large hub. In contrast, this hub is on the periphery closely related to "people", "citizens", "living" and "empowerment". The meaning of the empowerment present an analogy of team sport with standards, trainings, co-operation, engagement and further kind of confidence to mutual goals. This approach is significantly connected to the expected participation in the mentioned hub.

An unpredicted result is that "communities" has only one direct connection to the huge hub of participation, as well as, the category of "participation". It would be more connected as their functions to network of "people", collaboration or the imaginary "community". On the contrary, the expected participation is strongly tied to citizens or people as their living aspects. Moreover, "community" and "communities" are more interconnected separately with social-collective-public factors. Presenting the different role of the general term "community" and the functions of existing "communities", the connectivities are remarkable. Namely, the real communities are related to working social participation and an expected-imaginary "community" belongs to the holistic approach of smart city. Investigating with further content analysis, communities involve public values and public safety which are basic expectations of the participation. Participation is closely related to the aware citizens or to the private sector, and also, to organisations, institutes and fundings.

"Data" is an absolute centrum with close connectivity to both "private" and "public" sectors, and also, to "social" aspects and "people". This approach is in harmony with the smart city basic criteria (Ju *et al*, 2018), as well as with the most interconnected "open data" in the centre. However, the central large hubs are the "people" with their communities, and also, "social" or "public" factors. Additionally, "participation" is the peripheral hub with well-connected but less emphasised "empowerment" regarding collaborations and co-operations. "Collective" aspects and "collaboration" are medium-size nodes and only situated on the edge of the network. Consequently, the open data is mostly tied to public sector and social issues assuming the participation of people and their living aspects. Concerning this result and comparing it to the governmental outputs, the university research has more co-operation projects with the local governments than the local business.

Referring the first original research question, both individual and collective participation are expected in smart cities by smart citizens in context of current governmental-business-university research co-operations. Regarding the differences of subcorpora, the business has a strong interest in the collective and collaborative participation, while the government partly expects collective participation and partly the individual ones triggering civic experiments and community leadership, and the university research focuses mostly on the individuals with their "way of life" issues.

Studying the corpus based on the second research question, smartmentality presents a direct connection to the engagement and empowerment by citizens and local communities. Teamplayers are assumed in this contexts who are aware citizens and their communities, who trigger bottom up innovations and investments, who contribute to policy or design of the smart city. The following section is going to investigate further details of these results and analysing the most referred cities of the corpus.

## 5. Findings II. The most referred cities in the corpus

First and foremost, the most mentioned and cited cities have been filtered in the corpus. Although, different kind of city rankings are available in every year with various methodologies, this research study focused only on the corpus following the selected documents, research goals and research questions. In other words, there was not a goal to use any other rankings which are not comparable as their different methodologies.

The text network of the most frequently mentioned top 10 cities in the corpus has 29 nodes which is the sum of the 19 keywords and the top ten cities. This network has 319 edges which computes an average degree of 22 and the maximum number of degree is 28. The 1.22 average path length indicates the various hubs in the network as well as 2 diameter reflects words positioned away from these central hubs. The central hubs repeat the above summarised key issues, namely open data and the public-private sectors with strong ties where the public sector is the biggest and the one of the most connected hubs. The whole network is interconnected due to the 0.786 density. Groups of nodes, however, are more likely connected than it would be expected on the basis of chance in a random network. The strong tie of "people" or "citizens" and their data have the most significant connections, which confirms the previous results of keyword statistics. Although, the private issues present an interrelation to data and community, the community is a bigger hub. In conclusion, the collective participation is more expected than the individual ones (see Figure 4).

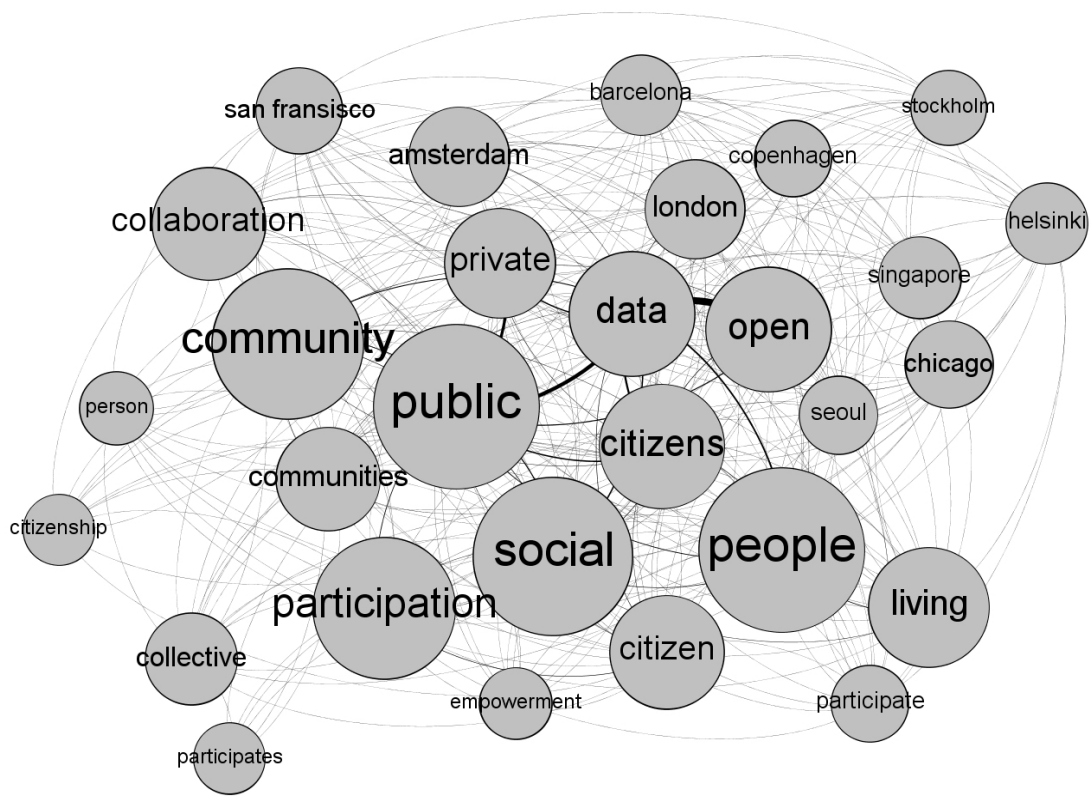

*Figure 4.* Text network of top ten cities

Amsterdam, Barcelona, Chicago, Copenhagen, Helsinki, London, San Fransisco, Singapore, Seoul and Stockholm represent the top ten cities. The intersections of the top ten cities are "public and open data" including "community" factors and "social" aspects. This result confirms again a focus on collective participation.
"Participation" is partly connected to the centre via "social-public" sectors and "communities", and partly to the margin via "collective" matters and the less highlighted "empowerment". Top ten cities expect collective participation but this expectation is only a part of the smart city strategy, not a cornerstone in itself. Moreover, the category of "participation" has only indirect connections to the top ten cities. In line with this, the smartmentality with contribution to policy making is less emphasised. Otherwise, the "participation" category is connected directly to huge central hubs, such as open data, public-private sectors, social and collective issues, communities or the empowerment. Consequently, the key issues and participation types are available in every strategies of top referred cities.

Most of the top cities are situated partly or exactly on periphery in the network. However, their majority presents significant interconnectivity with each other and their reference points. Illustrated with examples, Helsinki has remarkable path to "living" factor and further Nordic countries. San Fransisco and Amsterdam are strongly connected to each other and "collaboration" projects, while the position of Seoul is more related to "open data" with aspects of "citizens" and "people". These cases are representing cultural and social-economic similarities or differences among the cities and the mutual references to each other.

The most connected and centralised three cities are London, Amsterdam and Seoul. In contrast, the "citizen" or "community" integration to their city operation by "collaboration" and "empowerment" are less frequent according to the text network. Interpreting this result, the smartmentality is not a strong part of the expectations in strategy and practice. Only San Fransisco and Amsterdam have shorter path to "collaboration" and types of engagement to highlight the role of smartmentality.

The category "private" is presenting a semi-central position with strong interconnectivity to nodes of London, Barcelona, Amsterdam and San Fransisco. The reason behind this is probably the personal data regulations in the European Union or the role of data centres in the U.S. IT hubs.

Beyond the close reading of the content network of top ten cities, details of concepts reveal the diverse landscape of expected participation in smart cities. The next sections are going to provide a short a summary about this based on the strongly connected cities and their contextual hubs.

### 5.1. Participation in policy making, living labs, improved smartmentality

Starting with the most centralised and highlighted node, London presents a holistic concept in smart city and smart citizen context. According to the corpus, the Smart London Board supports representatives of authorities, academia and leading technology sectors with interactive communication to individual or organisational participants. Additionally, the platform-based London Data Store and the Open Innovation 2.0 program promotes an information value chain. The digital infrastructure belongs to user-centric services, living labs and applied research projects to recycle data and provide feedback by citizens to improve city services. Institutional and collective participation are facilitated by empowerment and by access to open data, allowing the collaboration in planning and operation of the city. Talk London project invites participation to policy making, online discussion, surveys and training programs to upgrade digital skills. The Future Cities Catapult supports services for disabilities, such as navigation program for blind citizens. Privacy matters are in the spotlight with strong focus on security, protection and prevention. The final goal of this strategy is to create the highest quality of life as possible.

### 5.2. Neighbourhoods, early adopters, engagement along communication technologies

Amsterdam and San Fransisco, as the most interconnected hubs to communities and collaboration, develop mobility services and explore neighbourhoods by open data and telecommunication systems. Strategy of Amsterdam interprets inhabitants as end users and urban contributors with various types of engagement. Citizens, organisations and the government have established together a testing ground for new concepts in collective thinking. Collaborative projects are provided by circular economy between public and private sectors to economic and social values, to liveability and creativity, and also, to healthy and prosperous life. Insight of emotions and desires become public instantly by highly informed citizens and communities. According to the self-definition, Amsterdam is a city of early adopters where well-educated and multilingual population is receptive to digital technologies and innovations to support sustainability. Special education programs are shaping the adaptive attitude towards the usage of telecommunication services and smart applications. San Fransisco provides access to reach real time data flow for comfort and for demand-responsive pricing. Smart phone applications of local news and intensive usage of social media support the data driven thinking of residents. Additionally, robust community interaction facilitate engagement in neighbourhoods by open data.

**5.3. Citizen-led experiments, joy of participation, cybersecurity**
Regarding the text network position of Barcelona, which is a bridge between the mentioned nodes of San Fransisco and Amsterdam, and also, directly connected to Nordic cities with close link to London, the smart city strategy facilitates the citizens and communities to be developers or producers in city life with empowerment. Individuals are interpreted as integrated part of the technology and the upcoming artificial intelligence in line with the literature review. Human and non-human actors are also involved in dynamics of projects of Barcelona. To share a brief example, FabLab community is a high-technically experienced community with early adopters for knowledge-sharing and smartmentality. Citizens are challenged translating innovations into action in their districts and neighbourhoods. These cities represent strong collaborations and overlapped fields of the above mentioned factors in the corpus.
The top Nordic cities, namely Copenhagen, Stockholm and Helsinki define citizens as openminded individuals for collaboration in the government supported initiatives in the corpus. In case of Copenhagen, open platforms are welcome to citizen-led innovations and the joy of participation. Every implementation is subordinated to quality of life. Data protection, ethical considerations, and also, increased accessibility for smart tourists or disabled people are mentioned. Helsinki supports the on-demand services to motivate citizen participation and to increase digital awareness. Adaptations, sustainability and resilience improve the bottom up collaborations and start-up business. The increasing number of young families supposes participation in civic activities and experiments. Needs of citizens shape data-driven decisions and collect best practices to follow. Stockholm aims the best quality of life and high level of cybersecurity. Small households, neighbourhoods and districts are involved in the smart and intelligent environments. New social meeting spots and testing environments facilitate the social and business innovations.

**5.4. Surveillance culture, transparency, anonymity**
Singapore and Chicago are situated on the ring of Nordic countries in the text network with open services and community focus. In case of Singapore, improved smart culture and engagement of ageing population are strategic guidelines. Human capital, education and remote health care services are mentioned with the highest frequency. The city strategy presents a decentralised concept using the "rainforest" metaphor to interpret the necessary multiple redundancies in services, open data circulation and interactive communication. Additionally, a surveillance technology system detects actions for a law-abiding society and overall society control. The live-work-play environments and the commercial activities stimulates participation, interaction and socialisation. Trainings and integration projects support smartmenatlity and talented inhabitants engage people into the ICT culture. The closely related Chicago creates meaningful open data and contents by smart applications to facilitate collaborations among citizens, civic technology and government. The goal is to

"make every community a smart community". Trainings, engagement programs, smart community benchmarking, transparency and civic hackers support the smartmentality. Last but not least, Seoul with direct connection to Singapore and Chicago emphasises the role of anonymity. Three arguments support this approach. First, convenience motivates people to join smart and intelligent services. Second, citizens and visitors of a city share their data in case of anonymity for keeping privacy. Third, citizens express their ideas concerning public or political issues without personal consequences. All three of them presuppose responsible participation in city life and the slogan "city of happy citizens and a city beloved by the world".

Considering the summarised strategies, a diverse landscape of smart cities has been revealed. From the focus on the governmental control by surveillance technology to the empowerment to policy making several approaches are available. Based on the content network, the top ten cities have strong ties not only to the key issues, such as open data, public and private issues or communities, but also to each other. The reason behind this output is the comparative methods of the documents, and also, the competitive context. Studying the highlighted and referred strategic elements, the data driven thinking facilitates the public and private sector, and also, communities and citizens.

Cultural, economic or social contexts influence deeply the presented approaches. To support this statement with illustrations, Seoul represents the cultural trend of South Korean internet anonymity, Singapore prefers the surveillance technology in the highly regulated society, Amsterdam and the Nordic cities emphasise on the democratic values and transparency, San Fransisco focuses on business issues as an ICT-hub. Obviously, the cultural-social-economic background requires further deep investigation in a next round.

Most of the top cities are definitely presuppose participation and empowerment regarding the research questions. Participation types are connected significantly to activities from data sharing to contribution of policy making primarily by technology promoters, early adopters and young families. The smartmentality is significantly connected to digital communication technologies causing constant engagement or empowerment. A thought-provoking result is the importance of emotional engagement and the happiness factor of individuals.

Beyond the content network and the answers for the research questions, further significant results have been also revealed. The universal and ultimate goals of the top ten cities are the high level life quality and wellbeing by economic and social values. The key to the smartmentality is the education. Only a few cases, however, are representing generation-specific developments, smart tourist programs or additional services in case of disabilities. Moreover, ethical questions are hidden frequently in the city documents, which implies a probably massive challenge for the cities in the context of upcoming AI.

## 6. Conclusion and recommendations

The paper has discussed expected participation and mentality of smart citizen in smart cities. The results confirmed the statement of Gudowsky and his co-authors (2017) about the fundamental role of human factor. Moreover, it was also verified that smart citizens are not only datasets (Johany and Bimonte, 2016) but also active or facilitated participants of the smart cities in policy making and collaborations.

Recalling the research questions, various expected participation types are available and smartmentality has diverse manifestations. Open data of inhabitants and adapted smart services define the ultimate expectations. In contrast, an unpredicted result is that the relevance of anonymity in open data systems appears only in one city strategy as a key concept. Collective participation is the most highlighted contrary to the individual one, primarily for business strategies and least for university research. Regarding the smartmentality, the collaborative engagement is partly supported by governmental and business strategies which confirms the role of proactive "participatory urbanism" (Han and Hawken 2018). The socio-technical engagement (Barns 2018) and the intensive usage of communication technology trigger bottom up innovations, investments or contributions to policy making via digital platforms. Additionally, the neighbourhood-based practice build smart communities and living labs to support the adaptations in local context. These results are confirming ideas from the literature review, primarily the significant role of participation in policy economy (Gudowsky 2017) and the key aspect of responsibility (Vanolo 2014) as a result of empowerment. The empowerment is less represented in the top down decision making but its options are emerging from micro-business to civic experiments.

Certain potentials are underrepresented or hidden in the corpus. First of all, smart tourism was mentioned but without specific concept and only in a few cases. However, scholarly articles highlight the role of visitors, not only the citizens (among others Lim *et al* 2018). Nevertheless, all the top ten cities from the corpus are definitely tourist destinations. Furthermore, it seems to have a luxury decision to hide or forget the role of expatriates or temporary residents in global multiculturality. The competitiveness of the cities, the image of liveability through internet media or the city rankings influenced by this group. Consequently, the collective participation perhaps should be segmented to local inhabitants, temporary residents, expatriates and tourists.

An additional crucial point is the lack of specific spotlight on privacy issues, ethical questions and consequences of the upcoming artificial intelligence existing in symbiosis with the human-nonhuman values. Without these considerations, smartmentality will exist only as a slogan slightly without meaningful content. The expected participation and responsibility-based smartmentality assume holistic approach. It is particularly valid if the ultimate goal is the the well-being and the liveability. The output depends not only on the local context (Neirotti *et al*, 2014), but also on moral and private issues. The upcoming technologies, such as AI from the literature review or the merged physical and the virtual scenes from the text networks, require the understanding and sensitisation for more sophisticated smartmentality.

Last but not least, metaphors in the corpus are presenting a relevant interpretative framework to understand the interrelation of smart city and smart citizen. The "rainforest" metaphor captures a redundant city operation with always available services and their changes. The "team player" metaphor interprets the citizen as an actor in a human-non human network to collaborate. The additional metaphor "public agora" from the literature review (Vanolo 2016) expand this approach with a public space as a centre for team players, early adopters or further functions from the corpus.

Investigating the final results, three key recommendations have been formulated. First, the forthcoming technology with structural changes and with AI should facilitate a strategy of human and non-human symbioses more than ever. A sophisticated approach with ethical and privacy issues are required for the higher level of smartmentality. Second, understanding key factors of the participation, such as local context, cultural differences, temporary or permanent human roles, neighbourhoods and communities, result in competitiveness and higher quality of life. Third, further metaphors are suggested to consider, interpret and communicate the interrelation of smart city and smart citizen. This recommendation is particularly valid for the cultural-social-economic perspective improving the upcoming structural-technological changes and reducing their risks.

**Acknowledgement**
This paper has been written with the support and within the framework of: KÖFOP 2.1.2 – VEKOP – 15-2016-00001 Public Service Development for Establishing Good Governance: The Digital Governance and Research Program.

**Sources**
Barns, S. (2018). Smart cities and urban data platforms: Designing interfaces for smart governance. *City, Culture and Society. 12*, 5-12. https://doi.org/10.1016/j.ccs.2017.09.006
Barns, S., Cosgrave, E., Acuto, M. & McNeill, D. (2017). Digital Infrastructures and Urban Governance. *Urban Policy and Research*, *35*(1), 20-31. https://doi.org/10.1080/08111146.2016.1235032
Bastian, M., Heymann, S., & Jacomy, M. (2009). Gephi: an open source software for exploring and manipulating networks. *ICWSM*, *8*, 361-362.
Benoit, G. & Hiroko, K. (2016). How are citizens involved in smart cities? Analysing citizen participation in Japanese "Smart Communities". *Information Polity*, *21*(1), 61-76. https://doi.org/10.3233/IP-150367
Blondel, V. D., Guillaume, J. L., Lambiotte, R., & Lefebvre, E. 2008. Fast unfolding of communities in large networks. *Journal of Statistical Mechanics: theory and experiment*, 10, 10008. https://doi.org/10.1088/1742-5468/2008/10/P10008
Brandes, U. (2001). A faster algorithm for betweenness centrality. *Journal of Mathematical Sociology*, *25*(2), 163-177. https://doi.org/10.1080/0022250X.2001.9990249
Brown, V. A. (2015). Utopian thinking and the collective mind: Beyond transdisciplinarity. *Futures, 65*(1), 209-216. https://doi.org/10.1016/j.futures.2014.11.004
Burgess, A. (2018) *The Executive Guide to Artificial Intelligence*. London: Palgrave Macmillan.
Danowski, J. A. (2013). *WORDij version 3.0: Semantic network analysis software*. Chicago: University of Illinois at Chicago.
D'Onofrio, S. & Portmann, E. (2017). Cognitive Computing in Smart Cities. *Informatik-Spektrum 40*(1) 46-57. https://doi.org/10.1007/s00287-016-1006-1
Feher, K. (2018) Contemporary Smart Cities: key issues and best practices. Berntzen, L. (Ed) SMART 2018: The Seventh International Conference on Smart Cities, Systems, Devices and Technologies, Barcelona, IARIA, 5-10.
Feher, K. (2017). Netframework and the digitalized-mediatized self. *Corvinus Journal of Sociology and Social Policy, 8*(1) 111-126. https://doi.org/10.14267/CJSSP.2017.01.06
Fortunato, S. (2010). Community detection in graphs. *Physics Reports*, *486*(3), 75-174. https://doi.org/10.1016/j.physrep.2009.11.002
Greenfield, A. (2013). *Against the smart city*. New York: Do projects. London, UK: Verso.
Gudowsky, N., Sotoudeh, M., Capari, L. & Wilfing, H. (2017) Transdisciplinary forward-looking agenda setting for age-friendly, human centered cities. *Futures, 90*(June), 16-30. https://doi.org/10.1016/j.futures.2017.05.005
Han, H. & Hawken, S. (2018). Introduction: Innovation and identity in next-generation smart cities. *City, Culture and Society, 12*(March) 1-4. https://doi.org/10.1016/j.ccs.2017.12.003
Hernafi, Yassine, Ahmed, Mohamed Ben and Bouhorma, Mohammed, "An approaches' based on intelligent transportation systems to dissect driver behavior and smart mobility in smart city", in. Proceedings of 4th IEEE International Colloquium on Information Science and Technology (CiSt), Morocco 24-26 October 2016.
Johany, F. & Bimonte, S. (2016). A Framework for Spatio Multidimensional Analysis Improved by Chorems: Application to Agricultural Data. In: Helfert, M., Holzinger, A., Belo, O. & Francalanci, C. Data Management Technologies and Applications. 4th International Conference, DATA 2015, July 20-22. Revised Selected Papers. Switzerland: Springer International Publishing.
Ju, J., Liu, L. & Feng, Y. (2018) Citizen-centered big data analysis-driven governance intelligence framework for smart cities. *Telecommunications Policy, 42*(10) 881-896. https://doi.org/10.1016/j.telpol.2018.01.003
Ishida, T. (2017). Digital City, Smart City and Beyond. In: WWW '17 Companion Proceedings of the 26th International Conference on World Wide Web Companion. Perth, Australia, April 03 - 07, 2017. pp. 1151-1152.

Isin, E., & Ruppert, E. (2015). *Being digital citizens*. Lanham, MD: Rowman & Littlefield.

Jacomy, M. (2009) Force-Atlas Graph Layout Algorithm. Retrived at: http://gephi.org/2011/forceatlas2-the-new-version-of-our-home-brew-layout/

Krippendorf, K. (2018). *Content Analysis: An Introduction to Its Methodology*. Fourth Ed. London: Sage.

Lim, C., Kim, K-J. & Maglio P. P. (2018). Smart cities with big data: Reference models, challenges, and considerations. *Cities, 82*(2018) 86-99. https://doi.org/10.1016/j.cities.2018.04.011

Meijer, A., & Bolívar, M. P. R. (2016). Governing the smart city: a review of the literature on smart urban governance. *International Review of Administrative Sciences*, *82*(2) 392-408. https://doi.org/10.1177/0020852314564308

Neirotti, P., De Marco, A., Cagliano, A. C., Mangano, G., & Scorrano, F. (2014). Current trends in Smart City initiatives: .Some stylised facts. *Cities*, *38*(June) 25-36. https://doi.org/10.1016/j.cities.2013.12.010

Newman, M. (2010). *Networks: An introduction*. Oxford University Press.

Paranyushkin, D. (2011). *Identifying the pathways for meaning circulation using text network analysis.* Berlin: Nodus Labs. Retrieved at: http://noduslabs. com/research/pathways-meaning-circulation-text-network-analysis.

Phdungsilp. A. (2011). Futures studies' backcasting method used for strategic sustainable city planning, *Futures 43*(7) 707-714. 10. https://doi.org/1016/j.futures.2011.05.012

Soto, J. A., C., Werner-Kytola, O., Jahn, M., Pullman, J., Bonino, D., Pastrone, C., & Spirito, M. (2015). Towards a Federation of Smart City Services. *Proceedings of International Conference on Recent Advances in Computer Systems - RACS 2016.* Hail: Saudi Arabia: Atlantis Press 163-168.

Thomas, V., Wang, D., Mullagh, L., & Dunn, N. (2016). Where's Wally? In Search of Citizen Perspectives on the Smart City. *Sustainability*, *8*(3) 1-13. https://doi.org/10.3390/su8030207

Vanolo, A. (2014) Smartmentality: The Smart City as Disciplinary Strategy. *Urban Studies, 51*(5) 883-898. https://doi.org/10.1177/0042098013494427

Vanolo, A. (2016). Is there anybody out there? The place and role of citizens in tomorrow's smart cities. *Futures*, *82*(September) 26-36. https://doi.org/10.1016/j.futures.2016.05.010.